\documentclass[12pt]{iopart}

\usepackage{iopams}

\usepackage[utf8]{inputenc}
\usepackage[T1]{fontenc}
\usepackage[separate-uncertainty,exponent-product = \cdot,
output-product = \cdot]{siunitx}

\usepackage{graphicx}

\usepackage{hyperref}
\hypersetup{colorlinks=true,urlcolor= blue,citecolor=blue,linkcolor=blue}

\usepackage[nameinlink,noabbrev,capitalize]{cleveref}

\begin{document}

\title[Atomistic analysis of Li migration in LATP]{Atomistic analysis of Li migration in {$\text{Li}_{1+x} \text{Al}_x \text{Ti}_{2-x} {(\text{P} \text{O}_4)}_3$} (LATP) solid electrolytes}

\author{Daniel Pfalzgraf, Daniel Mutter$^{*}$ and Daniel F. Urban}

\address{
    Fraunhofer IWM, W\"ohlerstr.\ 11, 79108 Freiburg, Germany\\
    $^*$Author to whom correspondence should be addressed\\
    }
\ead{\mailto{daniel.mutter@iwm.fraunhofer.de}}


\begin{abstract}
We examine the ionic migration of Li in LATP [$\text{Li}_{1+x} \text{Al}_x 
\text{Ti}_{2-x} {(\text{P} \text{O}_4)}_3$] solid electrolytes from an atomistic 
viewpoint by means of density functional theory calculations. We vary the Al 
content and investigate its effects on the 
crystal structure of LATP and on the migration energy landscape of interstitial 
Li ions. The energy profiles governing the Li diffusion are found 
to be systematically influenced by the position of Al ions in direct vicinity 
of the migration path, and we derive a simplified classification scheme of three 
universal energy profile shapes. The overall influence of the Al/Ti-ratio on the Li 
migration is analyzed by a separation into chemical and geometrical aspects. 
This work provides a solid basis for a resource-efficient computational 
examination of the ionic conductivity of Li in LATP with varying Al/Ti 
concentrations.

\end{abstract}


\maketitle

\section{Introduction}\label{sec:introduction}
Modern society is currently facing rapid developments regarding its energetic 
needs. The importance of energy storage solutions is ever-growing, both in 
production, with the focus on renewable energy sources, and in consumption, 
where electronic miniaturization demands novel technologies. For hand-held 
electronics, the introduction of Li-ion batteries, which generally contain 
liquid electrolytes (LEs), has been revolutionary. Yet, intrinsic problems 
hamper the improvement of this technology, such as chemical instabilities at the 
LE/electrode interfaces at high voltages, or the growth of Li dendrites, which 
can cause short-circuits \cite{Tarascon2001}. 

Batteries using solid state electrolytes (SSEs) are regarded as a promising 
alternative circumventing these issues. Most solids show ionic conductivities 
which are many orders of magnitude lower than those of state-of-the-art LEs, 
apparently making SSEs a naturally bad choice for electrolytes. However, some 
materials exhibit so-called super-ionic conductivity, which elevates them to the 
same level as LEs regarding their diffusive properties. In addition, SSEs 
exploit many natural advantages of a solid, such as mechanical stability, which 
largely suppresses dendrite growth \cite{Li2014,Fan2018,Zheng2018}.

In this research, we examine the compound class of Lithium Aluminium Titanium Phosphates (LATP),
\begin{equation}
    \centering
    \text{Li}_{1+x} \text{Al}_x \text{Ti}_{2-x} {(\text{P} \text{O}_4)}_3 ~~~~\text{with}~~~~ (0 \leq x \leq 2),
    \label{eq:LATP}
\end{equation}
which has been investigated for its ionic conduction properties since the late 
1980s \cite{Aono1990}. LATP has seen an increased research interest in recent 
years, both in simulation and experiment.

As compiled by Rossbach \textit{et al.} \cite{Rossbach2018}, LATP is a 
superionic conductor for certain compositions ($0.2 \leq x \leq 0.5$). This is 
partially attributed to a growing number of mobile charge carriers with 
increasing $x$: In LATP, when substituting tetravalent Ti($4+$) by trivalent 
Al($3+$), additional mobile Li($1+$) ions are introduced due to charge 
compensation. However, the increase in ionic conductivity due to the introduced 
charge carriers was found to be diminished by other factors at values of 
$x>0.5$. Furthermore, it was found experimentally that the quality of ionic 
conduction in LATP is highly dependent on the method of synthesis 
\cite{Breuer2015}. The reason for these peculiarities is subject of active 
research.

The diffusion pathways of Li in LATP are well-investigated \cite{Arbi2013, 
Arbi2015, Monchak2016, Case2020}. Two different migration mechanisms have been 
analyzed by atomistic simulations: a vacancy mechanism involving one Li ion, and 
an interstitial mechanism involving the correlated movement of three Li ions. 
Experimental data and various atomistic simulations 
suggest that the interstitial process dominates the superionic conduction in 
LATP and related compounds \cite{Lang2015,Epp2015,He2017,Kuo2019,Zhang2020}.

In this study, we apply density functional theory (DFT) calculations to 
systematically investigate the effect of Al on the interstitial migration 
process of Li in LATP. In particular, we consider three different aspects: 
(i) the variation of the structural parameters as function of the Al content, 
(ii) the influence of a specific local Al neighborhood on the Li migration 
barriers and energy profiles, (iii) the dependence of these profiles on the cell 
volume as determined by the averaged global Al/Ti concentration.

The paper is organized as follows. In \cref{sec:structure}, we give a 
description of relevant LATP properties: Migration pathways of Li are 
introduced, and general implications of substituting Ti by Al are sketched. In 
\cref{sec:computational}, we describe the computational methods used to generate 
the data. The data and its direct implications are 
presented in \cref{sec:results}, divided into three parts corresponding to 
the investigated aspects: The first part (\cref{sec:results.1}) presents the 
structural data as a function of Al content. In addition, the dependence of the 
formation energy of the interstitial Li configuration on the distance to the closest Al ion 
is examined. The second part (\cref{sec:results.2}) focuses on the energy profiles 
of the interstitial migration mechanism by investigating their dependence on the 
local Al/Ti neighborhood. In the third part (\cref{sec:results.3}), we evaluate 
the dependence of the migration energy profiles on the volume of the crystal, in order 
to decouple geometrical from chemical effects, and to determine the influence 
of the global Al content. The results are discussed in \cref{sec:discussion}, 
and \cref{sec:conclusion} summarizes the gained insights and gives an outlook on 
further research directions.

\section{Methods and Model}\label{sec:methodsandmodel}
    \subsection{Structure of LATP}\label{sec:structure}
    LATP is a structural derivate of Sodium Zirconium Phosphate (NZP), first 
described by Hagman and coworkers \cite{Hagman1968}, which belongs to a family 
of superionic conductors commonly referred to as NASICON. NZP crystallizes in a 
rhombohedral lattice of space group $R\overline{3}c$. Its characteristic 
structure of oxygen polyhedra, arranged in so-called ``lantern'' substructures, 
stabilizes a three-dimensional network of migration channels suitable for small 
cations. These structural characteristics of NZP are present in LATP as well. 
More structural details were described extensively in previous papers (e.g. 
\cite{Lang2015,Mutter2019}). Therefore this section mainly focuses on a description of 
the migration channels and the migration mechanisms of Li.

\begin{figure}[htbp]
    \centering
    \includegraphics[width=0.6\textwidth]{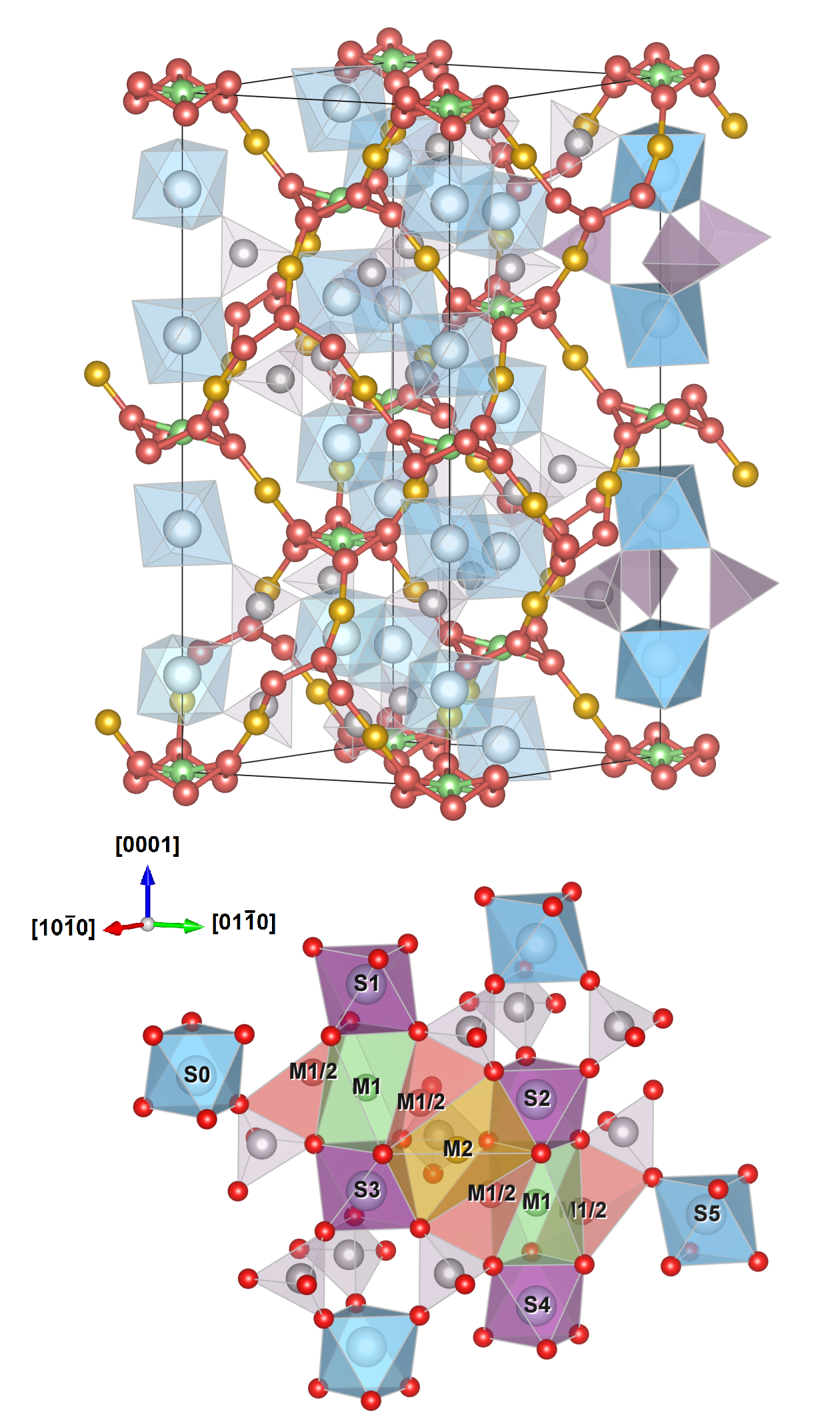}
    \caption{(Color online) Crystal structure of LATP. Top panel: hexagonal unit 
cell (enclosed by solid lines) with a schematic representation of the migration 
channels of Li ions. Light blue spheres represent sites occupied by Ti or Al 
ions, light grey spheres respresent P ions. The corners of the octahedra 
surrounding the sites of Ti and Al, and the tetrahedra surrounding P sites are 
occupied by O ions (not shown). Green, red brown, and yellow spheres connected 
by rods represent the M1, M1/2, and M2 sites of Li, respectively, as introduced 
in \cref{sec:structure}. Two characteristic ``lantern'' substructures, each 
consisting of two Al/Ti-octahedra (light blue) linked by three P-tetrahedra 
(light grey), are highlighted on the right. Bottom panel: detailed surroundings 
of a Li migration pathway. Oxygen ions are 
represented by red spheres. The labels S0--S5 denote specific sites of Al/Ti 
occupation (\textit{cf.} \cref{sec:results.2}). Polyhedra colored purple 
highlight the sites (S1, S2, S3, S4) for which occupations were varied in the 
calculations presented in \cref{sec:results.2}.} 
    \label{fig:migration_path}
\end{figure}

For our simulations, we choose a hexagonal unit cell as depicted in the top 
panel of \cref{fig:migration_path}. It contains 6 lantern units of stoichiometry 
$\text{Al}_x \text{Ti}_{2-x} {(\text{P} \text{O}_4)}_3$, two of which are 
highlighted on the right side of the figure. Interlinked lanterns form a rigid 
structure, encompassing connected cavities through which Li ions can migrate. As 
visualized by the balls-and-sticks framework in the top panel of 
\cref{fig:migration_path}, there exist certain discrete positions of Li ions 
being especially relevant for the diffusion processes, as described in the 
following.

\subsubsection{Li sites}
In LATP, there are three known 
symmetry positions of Li ions along the migration channels with special 
importance to the migration. They are shown in \cref{fig:migration_path} (bottom 
panel): On the ``M1'' position (Wyckoff site $6b$, colored green in the figure), 
the Li ion sits in between two neighboring lanterns along the $[0001]$ 
direction. In LTP ($x=0$, i.e. no Al ions in the structure), the occupation of M1 corresponds to the lowest energy of Li in the 
structure, meaning that in the ground-state, all M1 sites are occupied. The 
``M2'' site (Wyckoff $18e$, orange) is centered halfway in between two such M1 
positions. The Li migration mechanism in LATP occurs along this pathway defined 
by the M1 and M2 sites. 

When there are more Li ions in a cell than there are M1 sites, it is observed 
that an M1 site becomes vacant \cite{Rossbach2018}. DFT calculations show, that 
this occurs in favor of the occupation of two interstitial ``M1/2'' sites 
(Wyckoff $36f$, colored red brown in \cref{fig:migration_path}) in close 
proximity to the vacancy \cite{Lang2015}. For each M1 site, there are six 
possible M1/2 sites, arranged around M1 at slight tilt angles with respect to 
the $(0001)$ plane, in a three-fold rotation symmetry (\textit{cf.} 
\cref{fig:migration_path}). The occupation of two M1/2 sites directly 
opposite to each other with Li forms a stable M1/2 pair configuration. 

Note that the sites are not named consistently in the literature: In this paper, we use 
the notation and terminology of Lang \textit{et al.} \cite{Lang2015}.

\subsubsection{Migration mechanisms}

For low Li densities within the structural network, the main migration process for Li is a
\textit{vacancy mechanism} involving two M1 sites and their intermediary M2 site: 
A Li ion occupying an M1 site relocates to a neighboring vacant one, with an M2 
occupation as a transition state. This mechanism has been thoroughly examined via DFT
calculations and was found to have a migration barrier of $\Delta E\approx\SI{0.41}{\eV}$
in the case of LTP \cite{Lang2015}.

At higher Li densities, the \textit{interstitial mechanism} involving M1/2 pair 
occupations becomes relevant. In this case, a pair occupation relocates to a 
neighboring singly-occupied M1 site along the M2 site connecting them. One Li of 
the former pair then fully occupies the previously vacant M1 site, and the other 
forms a new M1/2 pair with the previously unpaired ion. In analogy to the 
vacancy process, we define the M2 transition state as the intermediate step of 
this process, where both M1 sites and the M2 site in between are occupied by one 
Li ion each. In LTP, this coincides with the energetic saddle point. 

From DFT simulations, the energy barrier $\Delta 
E\approx\SI{0.19}{\eV}$ of the interstitial mechanism in LTP 
was found to be considerably lower than that of the vacancy mechanism   
\cite{Lang2015}. However, this migration process results in an effectively 
reduced dimensionality of the network: The Li M1/2 pairs have a fixed 
orientation along which the migration process can occur, and transitions between 
the three possible orientations are suppressed by an additional energy 
barrier of the same order as that of the migration step itself \cite{Lang2015}. 
Still, the interstitial process is expected to be the main reason for the 
super-ionic conductivity of LATP \cite{Epp2015, He2017}. 

\subsubsection{Substitution of Ti by Al}
Due to its significantly lower barriers, the interstitial process could be 
facilitated when designing a superionic conductor based on the NZP structure, 
since the ionic conductivity scales as an inverse exponential of the migration 
barrier (Arrhenius equation). In LATP, as mentioned earlier, this can be 
achieved by increased substitution of Ti($4+$) by Al($3+$). The Al ion replaces 
a Ti ion on its site, leaving the general structure unchanged. An additional Li 
ion is required for maintaining charge neutrality, leading to an 
increase of the parameter $x$ (\textit{cf.} \cref{eq:LATP}). As the additional 
Li ions form M1/2 pairs with former M1 site ions, the interstitial 
process is expected to gain importance.
Experimentally, however, the benefits of this substitution to the ionic 
conductivity are limited: While the conductivity improves with increasing $x$ up 
to $x\approx0.5$, a decrease of both total and bulk ionic conductivity is 
observed for higher values of $x$ \cite{Aono1990,Arbi2013}. This decrease in 
total conductivity might be caused by grain boundaries and the formation of 
secondary phases, which is addressed in research dealing with improved synthesis 
methods \cite{Bucharsky2015,Bucharsky2016}. The reason for the decreased bulk 
conductivity is not yet understood.

In the following, we denote the Al occupation rate by the number of Al ions 
per hexagonal unit cell $n_\text{Al}$. In the case of
charge compensation by Li, the composition index $x$, introduced in \cref{eq:LATP}, is equal to $n_\text{Al}/6$. For the purpose of our simulations, we however 
decouple $n_\text{Al}$ from the number of Li ions per cell, $n_\text{Li}$, 
unless denoted otherwise.

    \subsection{Computational details}\label{sec:computational}
    The DFT calculations presented in this work were performed using the Quantum 
ESPRESSO (QE) PWscf and NEB routines \cite{QE-2009,QE-2017}. We used the 
ultrasoft Garrity-Bennet-Rabe-Vanderbilt (GBRV) pseudopotentials 
\cite{Garrity2014}, a plane wave basis with a cutoff energy of \SI{40}{Ry}, and 
the generalized gradient appropximation (GGA) of the exchange-correlation 
functional by Perdew \textit{et al.} (PBE) \cite{Perdew1997}. Brillouin zone 
integrations were carried out on a $\Gamma$-centered grid of $6 \times 6 \times 
2$ $k$-points following the scheme of Monkhorst and Pack \cite{Monkhorst1976}. 
Unless specified otherwise, each calculation was performed allowing for ionic 
relaxation using the Broyden-Fletcher-Goldfarb-Shanno (BFGS) algorithm 
implemented in the QE routines. Cell volume relaxations were carried out at fixed 
unit cell symmetry. Energy profiles and barriers were determined using the 
nudged elastic band (NEB) and climbing image (CI) methods \cite{Henkelman2000}. 
Sets of symmetrically inequivalent ionic configurations were constructed using 
the site-occupation disorder (SOD) code of Grau-Crespo \textit{et al.} 
\cite{Grau_Crespo_2007}. Crystal structures were visualized with \textit{VESTA} 
\cite{Momma2011}.

\section{Results}\label{sec:results}
    \subsection{Dependence of structural parameters and Li interstitial formation on Al content}\label{sec:results.1}
    \subsubsection{Cell parameters} \label{sec:cellparameters}
In order to examine the influence of the Al content in LATP on the cell 
parameters, we performed DFT calculations allowing ionic relaxation 
and optimization of lattice vectors for various configurations of the Ti/Al 
occupancy. Considering every possible occupation of the 12 Al/Ti sites in the 
LATP supercell (\textit{cf}. \cref{fig:migration_path}) by a combination of these 
elements yields in total 236 symmetrically inequivalent configurations. For 
$n_\text{Al} \in \{0,1,11,12\}$, there is only a single structure 
which needs to be considered. For $n_\text{Al}\in \{2,3,10\}$, we considered the 
full set of symmetrically inequivalent configurations, namely 9, 19 and 9, 
respectively. For $n_\text{Al} \in \{4,5,6,7,8,9\}$, there are $\{50, 66, 90, 
66, 50, 19\}$ configurations, respectively, out of which we randomly chose six 
each for further analysis. Independent of $n_\text{Al}$, the number of Li ions 
was kept constant at $n_\text{Li} = 6$.

\begin{figure}[htbp]
    \centering
    \includegraphics[width=0.8\textwidth]{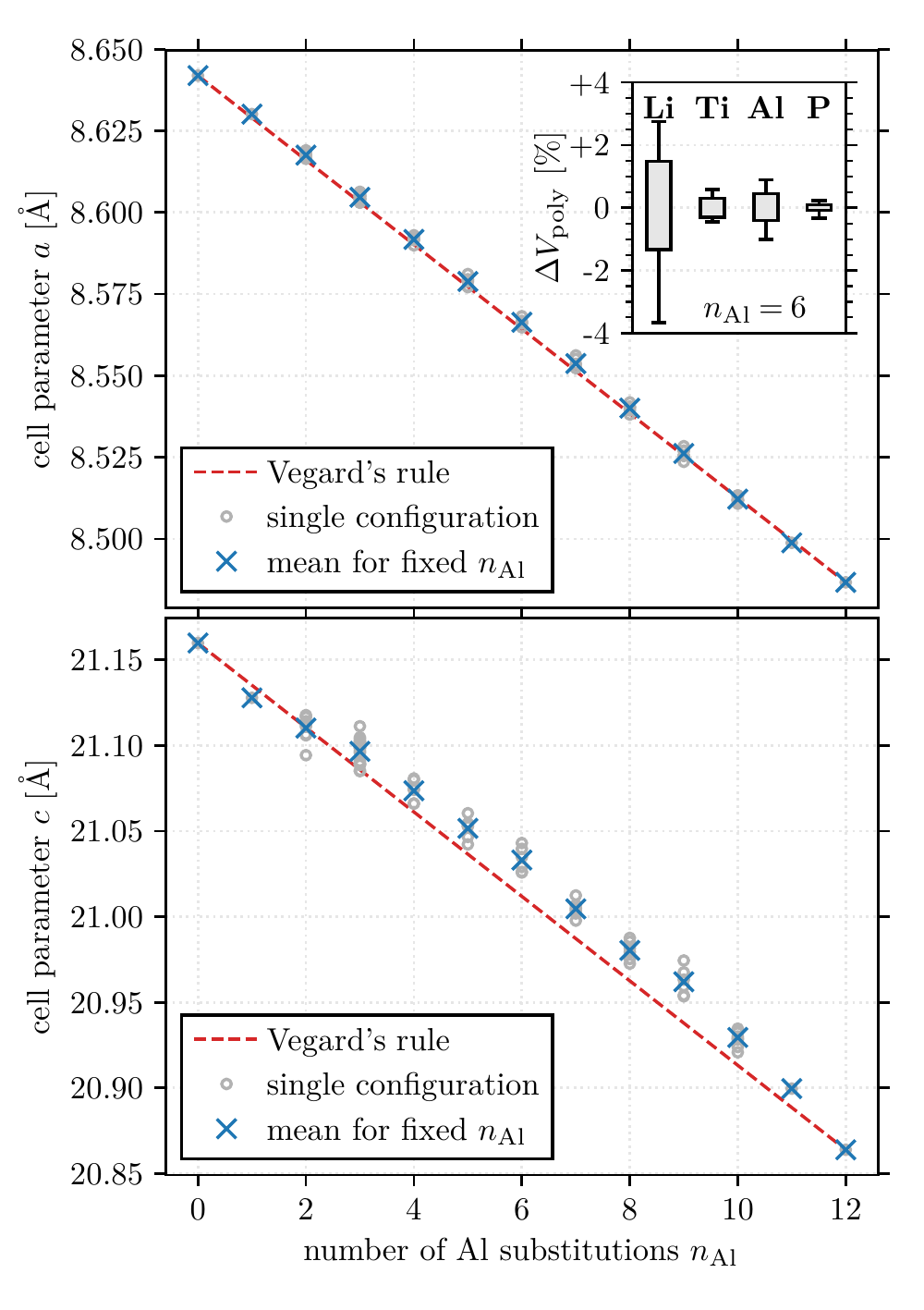}
    \caption{(Color online) 
Crystal lattice parameters of LATP as a function of the Al content. Upper 
panel: lattice parameter $a$, i.e. the length of the cell vector in 
$\langle01\overline{1}0\rangle$ direction. Lower pannel: lattice parameter $c$, 
i.e. the length of the cell vector in $\langle0001\rangle$ direction. The values 
obtained for the individual ionic configurations on the Al/Ti sublattice (gray 
circles) are averaged (blue crosses) for each number of Al ions in the 
simulation cell. Dashed red lines indicate a linear decrease as expected from 
Vegard's rule. In the inset, relative fluctuations of polyhedral volumes are 
visualized for the case of $n_\text{Al}=6$. For each element, the box represents 
the spread of all polyhedron volumes for all calculated cell configurations, 
relative to the mean of the element's data set. The vertical extent of the boxes 
encompass the $\SI{50}{\percent}$ of data points closest to the respective 
means, and the capped lines illustrate the most extreme volumes that occurred.}
    \label{fig:a_c}
\end{figure}

\Cref{fig:a_c} shows the resulting lattice parameters $a$ and $c$ corresponding 
to the cell dimensions in $\langle01\overline{1}0\rangle$ and 
$\langle0001\rangle$ directions, respectively. The cell parameter $a$ shows a 
linear decrease with the number of Al ions, closely following the 
empirical Vegard rule. For $c$, the deviation from the Vegard rule is more 
apparent, and individual configurations at the same $n_\text{Al}$ show a 
noticable spread. Upon increasing the Al content to its maximum, $a$ and $c$ decrease by $\SI{1.8}{\percent}$ and $\SI{1.4}{\percent}$, respectively. This 
leads to an increasing ratio $c/a$ with $n_\text{Al}$, implying an anisotropic 
cell volume change, as the cell contracts stronger in the $\{0001\}$ planes than along 
the $\langle0001\rangle$ directions.

\subsubsection{Polyhedra}

To further investigate the anisotropic change of the lattice parameters with 
$n_\text{Al}$, we analyzed the variation in the volumes of the polyhedral substructures.
To this end, the volumes of the oxygen polyhedra surrounding the 
cations were calculated for all examined configurations with $n_\text{Al}=6$. 
Large differences between the polyhedra are observed, as visualized in the inset 
of \cref{fig:a_c}. While the small P tetrahedra show only little fluctuations, 
the fluctuations become noticeably larger for the Ti and Al octahedra, and the 
volumes of the Li(M1) octahedra vary the most.

\subsubsection{M1/2 pair occupation}
\begin{figure}[htbp]
    \centering
    \includegraphics[width=0.8\textwidth]{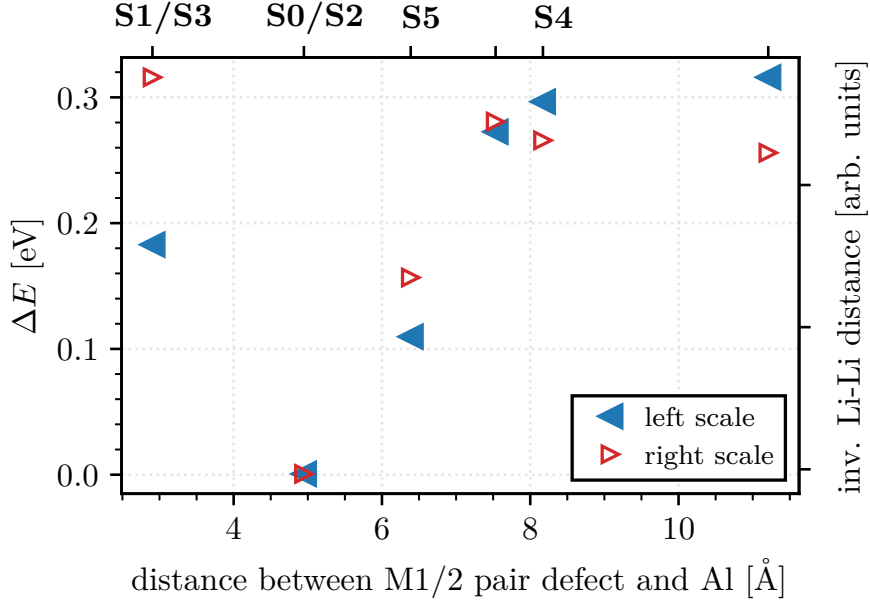}
    \caption{(Color online) Interaction between a Li M1/2 pair and a 
nearby Al ion. Filled, blue triangles pointing left: Energy ($\Delta E$) 
of one Li M1/2 pair in an LATP unit cell with $n_\text{Al} = 1$, with 
respect to the lowest energy of this data set, as a function of the distance to 
the closest Al ion. The position of the M1/2 pair is chosen as the 
geometric mean of the positions of its two constituting Li ions. Open, red 
triangles pointing right: Inverse distance between the two Li ions of the M1/2 
pair given in arbitrary units. Periodic boundary conditions are 
considered for all distance calculations. The labels on the upper horizontal 
axis relate the (M1/2-M1/2)-Al distances to the sites occupied by Al as 
visualized in \cref{fig:migration_path} (bottom panel), given that the M1/2 pair 
is located between S0-S1-S2-S3.}
    \label{fig:Al_occ_energies}
\end{figure}

In order to analyze the influence of the ionic structure on Li migration, we 
first considered a single Li M1/2 pair 
($n_\text{Li} = 7$) and a single Al ion ($n_\text{Al} = 1$) in our LATP 
supercell model. We calculated the total energies 
of all possible inequivalent configurations of this setup by relaxing the 
ionic positions while keeping the cell dimensions fixed at the values 
$a=\SI{8.63}{\angstrom}$ and $c=\SI{21.13}{\angstrom}$, obtained in 
\cref{sec:cellparameters} for $n_\text{Al}=1$. The resulting energies are 
plotted in \cref{fig:Al_occ_energies} with respect to the distance between the 
Al ion and the M1/2 pair. As a measure of the M1/2 pair position, 
we chose the mean of the position vectors of the two constituting Li ions. 
Energy differences of up to $\SI{0.32}{\eV}$ can be observed, and the 
non-uniform energy-distance relationship indicates anisotropic interaction 
between the Al ion and the Li M1/2 pair. Additionally, we 
calculated the inverse distances between the 2 Li ions on the opposing M1/2 positions, 
which are also plotted in \cref{fig:Al_occ_energies}. These distances show a 
behavior similar to the energy, suggesting a significant contribution of 
M1/2-M1/2 Coulomb repulsion to the interaction energy.

As a brief summary of \cref{sec:results.1}, it is observed that the variation of the Al 
occupation leads to cell volume changes in agreement with the Vegard rule. The polyhedra
around a given ion type in a given composition vary in size. Furthermore, it is seen that
the total energy of the Li M1/2 pair is sensitive to its relative position with respect to a nearby Al ion.

    \subsection{Influence of nearby Al ions on the Li migration energy profiles}\label{sec:results.2}

With the calculations presented in this section, we examine the effect of Al on 
the migration energy landscape of Li M1/2 pairs. To this end, we 
consider a set of four Al/Ti sites in the vicinity of the Li migration path, 
namely the sites S1, S2, S3 and S4, as displayed in \cref{fig:migration_path}, 
and vary their occupations. With all other sites in the cell being occupied by 
Ti, the resulting 10 symmetrically inequivalent configurations can be 
categorized into 4 symmetric configurations and 6 asymmetric configurations: The former are point symmetric around the M2 site 
with respect to their Ti and Al occupations, while the latter are not. For each 
of the 10 configurations, the energy profile between initial and final state of 
the interstitial mechanism (\textit{cf.} \cref{sec:structure}) was calculated 
with the NEB method. Again, only one Li M1/2 pair was placed in the cell 
($n_\text{Li}=7$). The cell parameters $a$ and $c$ were chosen according to the 
mean values of the respective Al contents $n_\text{Al}$, as derived in 
\cref{sec:results.1}. They were kept constant during the NEB calculations. For 
the calculations of the asymmetric configurations, 11 NEB images were used 
between initial and final state. For the symmetric configurations, only the 
first half of the path needs to be examined, as the other half is symmetrically 
identical. In this case, 7 intermediary images were calculated between inital 
and M2 transition state. For both sets, the climbing image (CI) method was 
applied to accurately sample the saddle point energy.

\begin{figure}[p]
    \centering
    \includegraphics[width=0.8\textwidth]{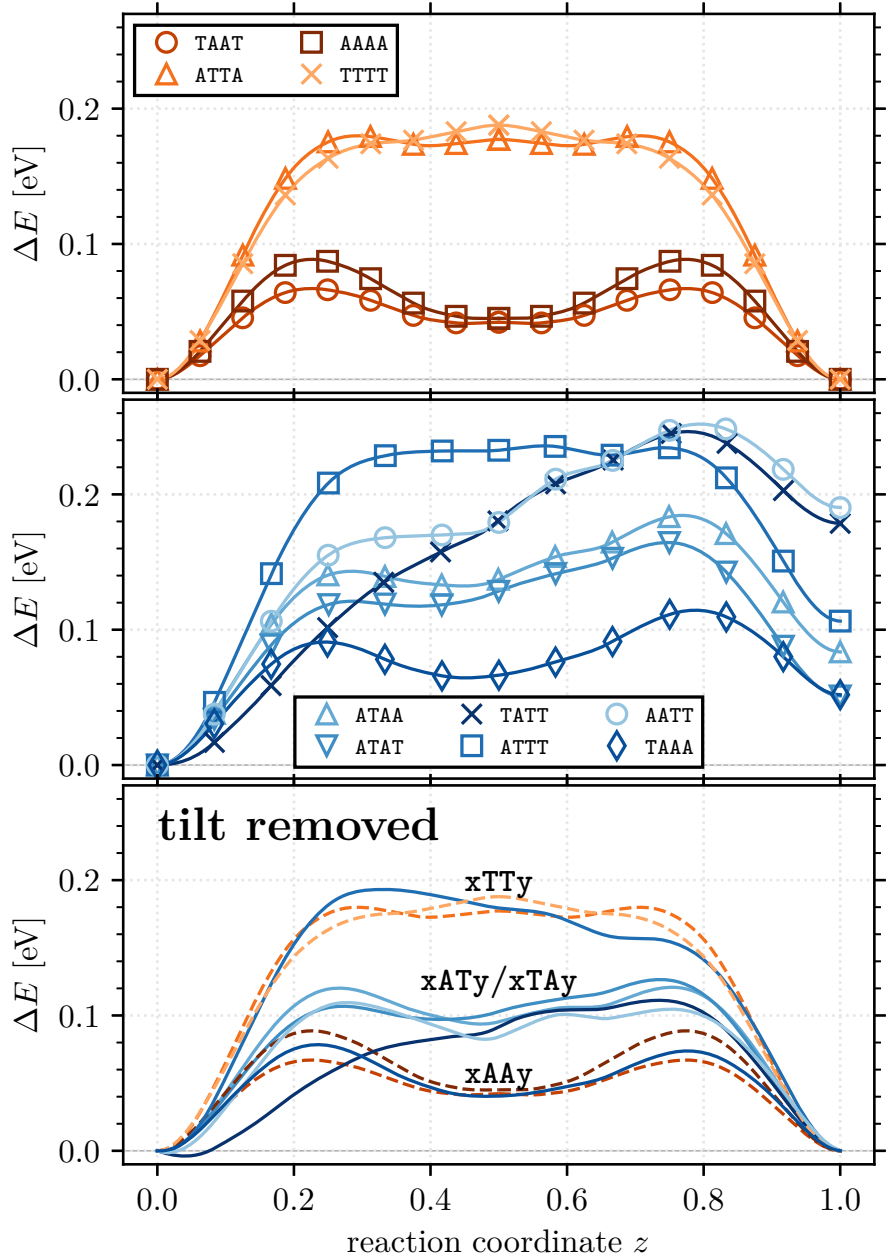}
    \caption{(Color online) Energy profiles of the interstitial migration paths in 
LATP, from initial ($z=0$) to final state ($z=1$), with respect to the energy of the former, for various neighboring 
Ti/Al-configurations. The top panel shows the symmetric profiles, i.e. those 
where initial and final state are identical, and the central panel shows the 
asymmetric profiles. The corresponding configurations are named as defined in 
\cref{sec:results.2}. Subtracting the energy slope between initial and final 
state yields the shapes displayed in the bottom panel, where dashed and 
continuous lines represent symmetric and asymmetric profiles, respectively.}
    \label{fig:barriers_shape}
\end{figure}

To analyze the data, we introduce a naming scheme addressing the occupational 
configurations and their corresponding energy profile by a string of four 
letters. It is constructed by taking the first letter of the elements occupying 
the Ti/Al sites (i.e. \texttt{A} for Al and \texttt{T} for Ti) in the order S1, 
S2, S3, S4 as enumerated in \cref{fig:migration_path}. In this way, symmetry and 
neighborhood relations are intuitively visible: A symmetric name corresponds to 
a symmetric energy profile (e.g. \texttt{ATTA}). Names that are mirrored 
versions of each other indicate equivalent configurations (e.g. \texttt{ATAA} 
and \texttt{AATA}), and the first (last) three letters denote the occupations 
surrounding one (the other) M1 site. The letters \texttt{x} and \texttt{y} 
denote arbitrary Ti/Al occupations, and names containing \texttt{x} and 
\texttt{y} reference the set of possible configurations.

The energy profiles of the symmetric configurations, with respect to the energy of their initial states, are presented in the top 
panel of \cref{fig:barriers_shape}. A clear distinction between two different 
profile types is apparent: The two paths with Ti closest to the M2 position, 
\texttt{TTTT} and \texttt{ATTA}, yield quite similar energy profiles along the 
migration path, with an energy barrier of around \SI{0.18}{\eV}. The M2 
transition state is a global energy maximum for \texttt{TTTT}, and a local 
one for \texttt{ATTA}. The main difference of the two curves is the presence 
of an energy maximum between initial state and transition state for 
\texttt{ATTA}, in contrast to the \texttt{TTTT} curve, which increases 
monotonically in this interval. The curves for \texttt{TAAT} and \texttt{AAAA} 
differ significantly from those for \texttt{TTTT} and \texttt{ATTA}, both in 
height and shape. Most notably, their energy barriers are reduced by 
about \SI{0.1}{\eV}, to \SI{0.07}{\eV} (\texttt{TAAT}) and \SI{0.09}{\eV} 
(\texttt{AAAA}). Moreover, the M2 transition state is a local energy minimum 
surrounded by two maxima.

The profiles of the asymmetric paths (\cref{fig:barriers_shape}, central panel) 
are more diverse at first sight. The energy differences between initial and 
final state vary at a scale of the order of the barriers themselves. They are in 
agreement with the Li M1/2 pair formation energies shown in 
\cref{fig:Al_occ_energies}. For example, the initial state of configuration 
\texttt{TATT} has an Al occupation on an S2 type site. The final state of the 
same configuration, equivalent to the initial state of \texttt{TTAT}, has an S3 
site occupied by Al. Both in \cref{fig:Al_occ_energies} and 
\cref{fig:barriers_shape}, one finds an energy difference of about 
$\SI{0.18}{\eV}$ between these states.

Despite the energy differences between initial and final states, the shapes of 
the asymmetric curves show similarities to the symmetric ones. For a better 
comparability, we remove the tilt in the curves by subtracting a constant 
gradient. This corresponds to a decomposition of the energy profile along the 
reaction coordinate, $E(z)$, into a slope term $E_\text{slope}(z) \propto z$ and 
a shape term $E_\text{shape}(z)$:
\begin{eqnarray}
    \centering
    E(z) &= E_\text{slope}(z)+ E_\text{shape}(z)\nonumber\\
    &= \left[E(1)-E(0)\right] \cdot z + E_\text{shape}(z).
\end{eqnarray}
The resulting functions $E_\text{shape}(z)$ are depicted in the lower panel of 
\cref{fig:barriers_shape}, together with the symmetric curves for comparison. A 
split into three distinct types is observed: asymmetric profiles corresponding to paths where the Al/Ti 
positions neighboring the M2 site (types S2 and S3, \textit{cf.} 
\cref{fig:migration_path}) are both occupied by Ti or both occupied by Al, fall 
into the same categories as the respective symmetric profiles. The set 
\texttt{xAAy} is characterized by a low barrier and a local minimum, and the set 
\texttt{xTTy} by a higher barrier and an energy plateau around the M2 
transition state. The other set of paths, $\texttt{xATy}$ and $\texttt{xTAy}$, 
forms a category of its own, with intermediate barriers.

As a brief summary of \cref{sec:results.2}, the calculated profiles show a great 
variance in their energy barriers and energy differences between inital and final states, 
hinting at a quite heterogeneous energy landscape for the migration of Li M1/2
pairs. However, the energy profiles have characteristic shapes, 
dictated by the Al/Ti occupation of the sites closest to the M2 site, allowing 
the determination of three distinct categories.

    \subsection{Influence of the cell volume on migration energy profiles}\label{sec:results.3}
    In the previous section, we examined the influence of local Al occupations 
within the unit cell on the migration path profiles. With a final set of calculations, 
we now investigate two further aspects: Primarily, we intend to decouple the 
chemical, electronic influence of nearby Al/Ti occupations from a purely 
geometrical one, caused by a deformation of the network of 
interconnected polyhedra (\textit{cf}. \cref{fig:migration_path}). Secondly, we aim to 
estimate the long-range influences of Al ocupations, conveyed through such 
mechanical deformations.

\begin{figure}[htbp]
    \centering
    \includegraphics[width=0.8\textwidth]{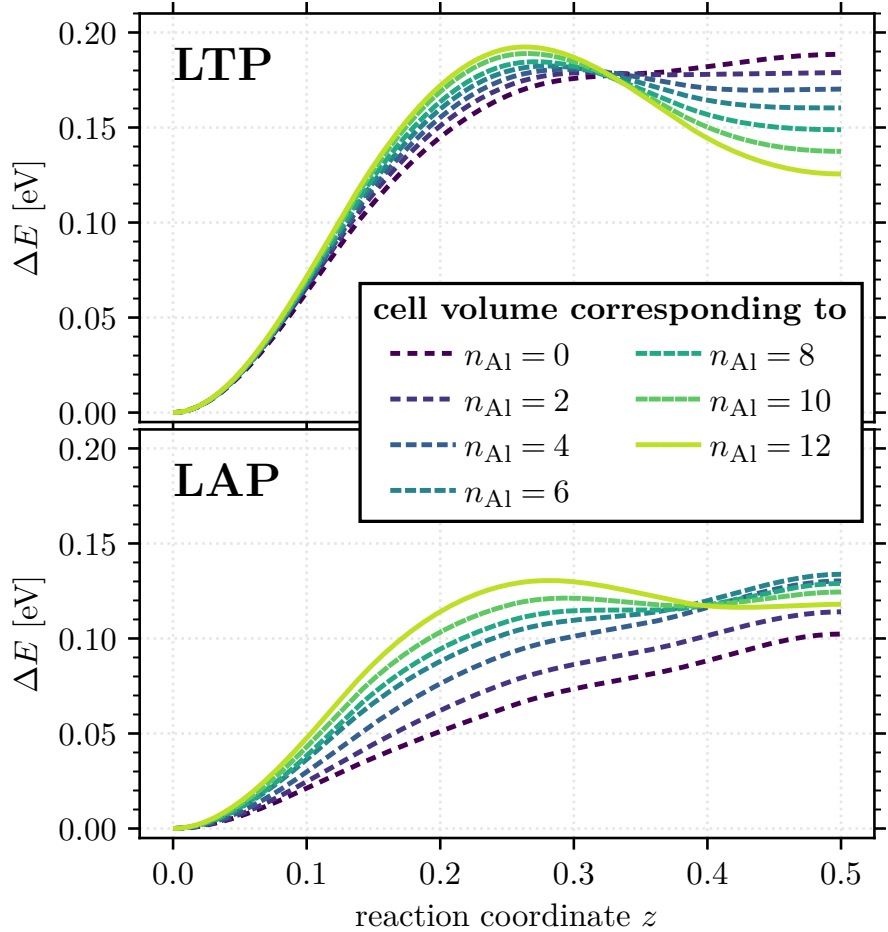}
    \caption{(Color online) Energy profiles of the interstitial migration path 
between initial ($z=0$) and M2 transition state ($z=0.5$) as a function of the cell volume for LTP 
(upper panel) and LAP (lower panel). The different curves were obtained 
by varying the cell volume. Therefore, the lattice parameters were chosen 
according to their averaged values at fixed Al content $n_\text{Al}$ (\textit{cf}. 
legend) as obtained in \cref{sec:results.1}. Darker line shades and wider separation of dashes indicate larger volumes.
}
    \label{fig:barriers_volume}
\end{figure}

Both of these problems can be tackled simultaneously by analyzing the influence 
of the unit cell volume on the energy profiles. In order to compare the 
different behavior caused by either a Ti or an Al occupation, we examine the 
two limiting cases: (i) pure LTP (no Al ions, $n_\text{Al}=0$) and (ii) the 
fully substituted simulation cell (all Ti ions replaced by Al, 
$n_\text{Al}=12$, ``LAP''). In both cases we consider one Li M1/2 pair 
occupation (i.e. $n_\text{Li}=7$). We carried 
out seven NEB migration path calculations for each of the two  
compounds, each at a constant, different volume. These volumes were determined by the cell 
parameters $(a,c)$ corresponding to their mean values for $n_\text{Al} \in 
\{0,2,4,6,8,10,12\}$ as obtained in \cref{sec:results.1}. As the paths are 
symmetrical for both compounds, the energy landscapes are computed between 
initial and transition state only, and discretized by five intermediary images. 
The resulting profiles are depicted in \cref{fig:barriers_volume}. 

In the case of LTP, a clear transition between profile shapes is visible. For 
large volumes (corresponding e.g. to a small average $n_\text{Al}$ in an 
extended material), the shapes are in agreement with those of type \texttt{xTTy} 
discussed in \cref{sec:results.2}: There is an energy plateau in the proximity 
of the transition state, which is a global maximum or a very flat minimum. For 
decreasing cell volumes (corresponding e.g. to larger average $n_\text{Al}$ 
values in an extended material) the transition state energy decreases and the 
intermediate maximum increases relative to the initial state energy, making the 
minimum of the transition state more pronounced. The overall energy barrier 
remains similar for all volumes. For the profiles at full Al occupation 
(LAP), a similar trend is visible at small volumes ($n_\text{Al} > 7$). However, 
for larger volumes corresponding to $n_\text{Al} \leq 6$, the intermediate 
energy maximum vanishes, and the relative transition state energy also lowers at 
a similar rate, resulting in a monotonic energy increase between initial and 
transition state at large volumes. 

\section{Discussion}\label{sec:discussion}

In this section, we discuss the results presented in 
\cref{sec:results.1,sec:results.2,sec:results.3} dealing with the 
different aspects of varying the Al/Ti occupation in LATP, 
which concern the crystal structure and the migration properties of Li.

In \cref{sec:results.1}, we described an anisotropic contraction of the LATP 
cell with increasing Al content and a fixed number of Li ions: With a relative 
change of $-\SI{1.4}{\percent}$ when going from LTP to LAP, the lattice 
parameter $c$ turned out to be less affected by the Al/Ti occupations compared 
to the lattice parameter $a$, which exhibits a relative decrease of 
$-\SI{1.8}{\percent}$. This yields an increase of the $c/a$ ratio by 
$+\SI{0.4}{\percent}$ between LTP and LAP. Individually, $a$ and $c$ shrink with 
increasing Al content following Vegard's rule, since six-fold coordinated Ti(4+) 
ions have an ionic radius of $\SI{0.61}{\angstrom}$ compared to 
$\SI{0.54}{\angstrom}$ of Al(3+) in the same configuration \cite{Shannon1976}. 
The different influence on $a$ and $c$ can be explained by the changes of oxygen 
polyhedron volumes described in \cref{sec:results.1} (\textit{cf.} inset in 
\cref{fig:a_c}): The polyhedra around Al, Ti and P are quite rigid due to the 
relatively high number of valence electrons of these elements, leading to strong 
bonds with the surrounding oxygen ions. Therefore, these polyhedra propagate 
local volume changes induced by introducing additional Al easily, by 
pushing and pulling neighboring polyhedra. The Li-M1-octahedra, however, are 
much more flexible, due to only a single valence electron of Li bonding with six 
oxygen ions. As the Al/Ti octahedra and the Li(M1) octahedra are aligned in 
$[0001]$-direction, the latter effectively act as a buffer, compensating local 
volume changes due to varying Al/Ti occupations by expansion and contraction in 
$[0001]$-direction, thereby lowering the impact on the overall $c$ parameter. 
This behavior was also described for LTP compounds with full substitutions on 
the Ti and P lattices \cite{Mutter2019}. Experimental data for LATP with $0 \leq 
x \leq 0.5$ (\textit{cf.} \cref{eq:LATP}) measured by Redhammer \textit{et al.} 
\cite{Redhammer2016} show an opposite trend: While $a$ decreases similarly to 
our results, $c$ shrinks notably more, resulting in a decreasing $c/a$ ratio 
with increasing $x$. In this case, however, the number of Li ions also changes 
with $x$, leading to a growing number of paired M1/2 site occupations, which 
were not considered in our calculations. Each of these occupations replaces one 
flexible Li(M1) octahedron by two stiffer Li(M1/2) tetrahedra, with stronger bonds 
between Li and O. Consequently, the overall buffering effect is weakened up to 
the point where the $c/a$ ratio decreases with $x$, which explains the 
discrepancy between the experimental data and our presented results. 
Nevertheless, the absolute volume change of the cell is rather small in both 
cases, implying high mechanical stability. Note that the cell parameters 
presented in this research systematically overestimate the experimental values 
by about $\SI{1.5}{\percent}$, as typically observed in DFT calculations using 
the PBE functional.

The nontrivial relationship between the energy of a Li M1/2 pair and its 
distance to the closest Al ion shown in \cref{fig:Al_occ_energies} can also be 
attributed to the large influence of structural effects in LATP. Direct 
Coulombic repulsion of the paired Li ions on M1/2 sites has a considerable 
impact on the formation energy of the pair, as its correlation with 
the inverse M1/2-M1/2 distance suggests. This distance is influenced by the 
position of the Al ion relative to the Li pair. In order to illustrate these 
relationships, we consider one M1/2 pair surrounded by the sites S0-S1-S2-S3 as 
defined in \cref{fig:migration_path}. Substitution of Ti by Al on an S0 (S2) 
site reduces the respective octahedron volume, pulling the M1/2 ions apart, 
which results in a reduction of the potential energy. Substitution of Ti 
by Al on the S5 position has a weaker influence, since it only indirectly 
acts on the M1/2 pair distance by increasing the space between neighboring 
lanterns. The more the Al ion and the M1/2 pair are separated, the smaller is 
the effect of the octahedron volume reduction on the M1/2 pair distance. The 
M1/2 pair distance vector is largely aligned in the $(0001)$ plane, i.e. it only 
has a comparatively small component in $[0001]$ direction. Consequently, a 
volume reduction of octahedra aligned in $[0001]$-direction, as e.g. induced by 
substitution of Ti by Al on an S1 (S3) site, effectively decreases the pair 
distance. This effect, which is strongest for an M1/2 pair directly adjacent to 
Al, leads to an increase of the potential energy. Coulomb repulsion between Li 
ions was reported by He \textit{et al.} \cite{He2017} to have a major influence 
on Li pair mobility in super-ionic conductors. Our findings of a correlation 
between M1/2 pair energy and inverse Li distance qualitatively support this 
statement, since the migration energy landscape, which essentially affects the 
ionic mobility, is to a large extent determined by the site energies of the 
migrating species. The considerable energy differences between differently 
surrounded M1/2 pair occupations of up to $\SI{0.32}{\eV}$, being of the order of 
the migration barriers, hint at a significant reduction of M1/2 mobility, as 
will be discussed in more detail below. Note however, that the calculated energy 
differences are likely to be overestimated: Due to the finite size limitations 
of the simulation cell considered in this study, the interactions 
between occupations in neighboring periodic cells are stronger in the $(0001)$-plane 
than in the $[0001]$ direction, since $a < c$. For larger simulation cells, one 
should expect a clearer convergence of the formation energies with distance for the 
long-range interactions.

The migration profiles presented in \cref{sec:results.2} exhibit considerable 
variance in shape and height, implying a diverse energy landscape for 
interstitial Li migration within LATP. In this regard, our research supports the 
assumption of Epp \textit{et al.} \cite{Epp2015} of a heterogeneous diffusion 
landscape based on measurements of Li diffusion activation energies. This also 
offers a possible explanation for the reported sensitivity of LATP bulk ionic 
conductivity to the synthesis parameters \cite{Breuer2015}, as an inhomogeneous 
spread of Al occupations could notably influence the overall connectivity of 
Li sites in the network. 

While Lang \textit{et al.} \cite{Lang2015} reported a significant energy 
difference between the M1/2 pair occupation and the M2 transition state, recent 
calculations by Zhang \textit{et al.} \cite{Zhang2020} suggest that these states 
should be nearly equal in energy. Our results cannot support the latter, but 
rather are consistent with the former observation.

The classification of the energy profiles into three categories shows the importance of 
the Al/Ti-occupation of the sites directly adjacent to the octahedron around the 
M2 position (sites S2 and S3 in \cref{fig:migration_path}) for the Li pathway. 
An increased Al occupation of these sites results in more favorable, i.e. lower, 
migration barriers. While the M2 transition state thereby tends to be 
stabilized, the large energy slopes, and the M1/2 rotational barriers which are
not investigated here, remain potentially significant obstacles for long-range 
Li diffusion. 
The importance of Al/Ti sites on the same $(0001)$-plane as the M1/2 pair occupation for the energetics, as discussed above, suggests that considering a larger environment of Al/Ti sites around the migration path may reveal an even finer categorization.

The calculations at varied cell parameters described in \cref{sec:results.3} 
show a transition between single-peak and double-peak energy profiles for Li 
migration as a function of the cell volume. It is therefore reasonable to assume 
that the change in profile shapes described in \cref{sec:results.2} can to some 
extent be attributed to the geometrical effect of a changed cell volume. As 
the double-peak occurs when a Li ion crosses the polyhedral face of the M2 
cavity, Li-O interactions are likely to be responsible for this characteristic shape.

\section{Conclusion}\label{sec:conclusion}
In this study, we analyzed the effect of the Al content and the ionic
distribution on the Al/Ti sublattice on the structure and Li ion 
migration properties of LATP super-ionic conductors. A mechanical picture of 
interlinked oxygen polyhedra, and their flexibility with respect to local 
geometrical changes, provides qualitative insights into the system's reaction 
to varying Al content and internal volume changes. Energy profiles for interstitial 
migration of Li M1/2 pairs can be categorized into three basic 
shapes apart from an overall energy slope. The characteristic profile shapes 
belonging to the different categories were shown to depend on the overall volume 
of the system, and the most favourable migration path occurs when two Al 
polyhedra are adjacent to the M2 cavity. The formation energies of Li 
M1/2 pairs, which determine the energy slopes between initial and final 
states of Li migration, vary on the scale of the barrier energies depending on 
their local ionic environment. These formation energies were found to strongly 
depend on the configuration of the surrounding Al/Ti site occupations, which 
defines the Li-Li distance of the M1/2 pair. A complete picture of the total 
influence of Al on the ionic conductivity of LATP requires the consideration of 
additional effects such as the existence of rotational barriers of M1/2 pairs, 
as well as the interactions between individual M1/2 pair occupations. 
However, the data presented in this paper serves as a solid basis to study Li 
diffusion in LATP on a larger scale using simplified models, e.g. by performing 
Kinetic Lattice Monte Carlo simulations for a network of Li sites in LATP 
with varying Al/Ti composition.

\ack \label{sec:acknowledgements}
This work was funded by the German Research Foundation (DFG), Grant No. El 155/26-1. The calculations were carried out on the computing facility ForHLR I of the Steinbuch Centre for Computing (SCC) of the Karlsruhe Institute of Technology (KIT), funded by the Ministry of Science, Research, and Arts Baden-W\"{u}rttemberg, Germany, and by the DFG.

\section*{References}
\nocite{*}

\bibliographystyle{iopart-num}
\bibliography{bibliography.bib}

\end{document}